\documentclass[epj]{svjour}
\usepackage{amsfonts}
\usepackage{graphicx}
\newcommand{\spinhalf}{\mbox{spin-$\frac12$} }

\begin{document}
\title{Exactly solvable model of topological insulator realized on \spinhalf lattice}
\author{Igor N.Karnaukhov \and Igor O.Slieptsov}
\institute{G. V. Kurdyumov Institute for Metal Physics, National Academy of Sciences of Ukraine, Vernadsky Ave. 36, Kyiv 03680, Ukraine}
\abstract {
In this paper we propose an exactly solvable model of a topological insulator defined on a \spinhalf square decorated lattice.
Itinerant fermions defined in the framework of the Haldane model interact via the Kitaev interaction with
\spinhalf Kitaev sublattice. The presented model, whose ground state is a non-trivial topological phase, is solved exactly.
We have found out that various phase transitions without gap closing at the topological phase transition point outline the separate states with different topological numbers.
We provide a detailed analysis of the model's ground-state phase diagram and demonstrate how quantum
phase transitions between topological states arise. We have found that the states with both the same and different topological numbers
are all separated by the quantum phase transition without gap closing. The transition between topological phases is accompanied by a rearrangement of the spin subsystem's spectrum from band to flat-band states.
}
\date{23.05.2014}
\maketitle

\section{Introduction}

An interesting new class of complex topological systems that include interacting spin and fermion sublattices
is presented and considered in the framework of the 2D exactly solvable model.
In contrast to traditional topological insulators, in this paper we consider a topological insulator interacting with spin sublattice.
Materials with strong correlation between electronic states of narrow (localized $f$) and conduction ($p,d$) bands
exhibit the heavy fermion behavior with exotic ground states~\cite{rev}.
The heavy fermion behavior mostly takes place in
the rare-earth based compounds, where localized-flat $f$ bands are strongly hybridized with $d$ conduction bands.
Recent experiments on $SmB_6$~\cite{ex1,ex2,ex3,ex4,ex5} have shown that this material belongs to this family.
$SmB_6$ is a strongly correlated topological insulator with dominating interaction between itinerant and narrow band electrons.

Alexandrov, Dzero and Coleman~\cite{t1} (also see~\cite{t2,t3}) assume that a state of topological insulator is formed in such materials due to the strong-spin orbit coupling and hybridization of the $\Gamma_8$ spin quartet of \emph{4f}-states with the \emph{d}-states of a conduction sea.
The momentum dependence of the hybridization amplitudes originates from the strong spin-orbit coupling in \emph{f}-bands. As a result, the family of heavy  fermion insulators (the Kondo insulators) are $\mathbb{Z}_2$ topological insulators.
We consider the model of the topological Kondo insulator with a Hamiltonian that breaks time reversal symmetry.

The topological phases in topological insulators~\cite{Hal,km,bhz,td,1,2,3}  and  spin systems~\cite{Kitaev,4a}
are characterized by  the Chern number. In the present paper we study a family of the models of complex systems in which the spin subsystem interacts with the fermion subsystem in a topological insulator state.
A non-trivial topological state is realized in the subsystem of spinless fermions, namely the phase of Chern topological insulator, when both inversion symmetry and time-reversal symmetry are simultaneously broken.
As far as we know, this is the first proposed exactly solvable model of a magnetic system whose exact ground state is a charge topological phase. Traditionally, a topological phase transition occurs between phases with different topological indices and the gap closing at a point of the phase transition is necessary in general.
Our purpose is to clarify the meaning of the quantum phase transition that is realized between phases with both the same and different values of the Chern number without a low energy gap closing.

\section{Model}

\begin{figure}[thbp]
\centering{\leavevmode}
  \includegraphics[width=8cm]{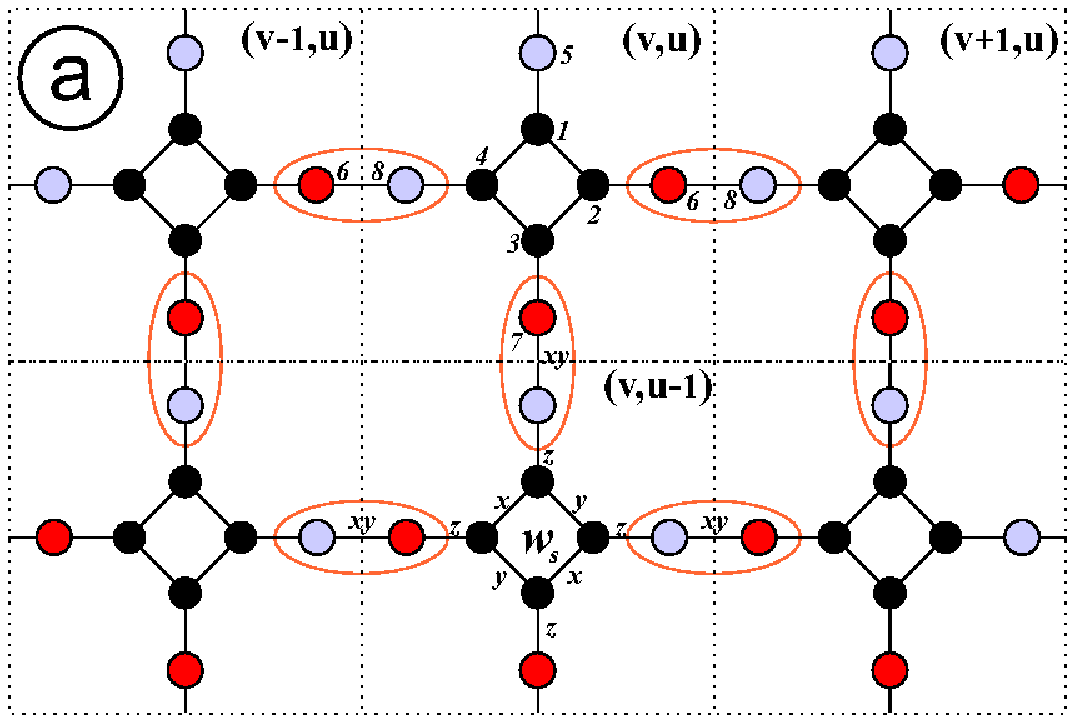}
  \includegraphics[width=6cm]{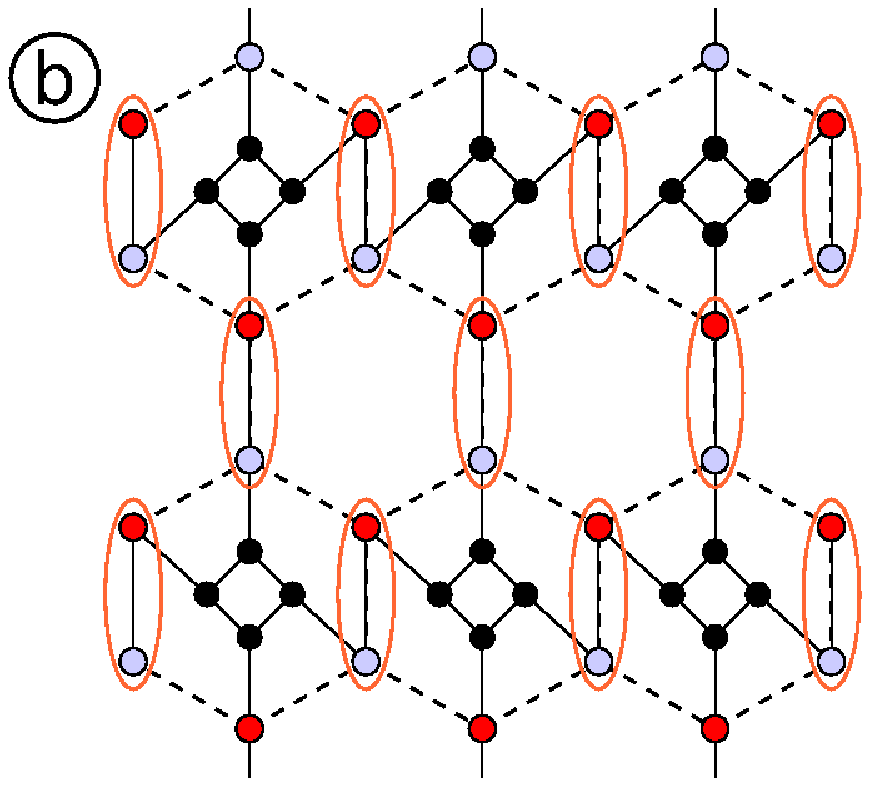}
  \includegraphics[width=6cm]{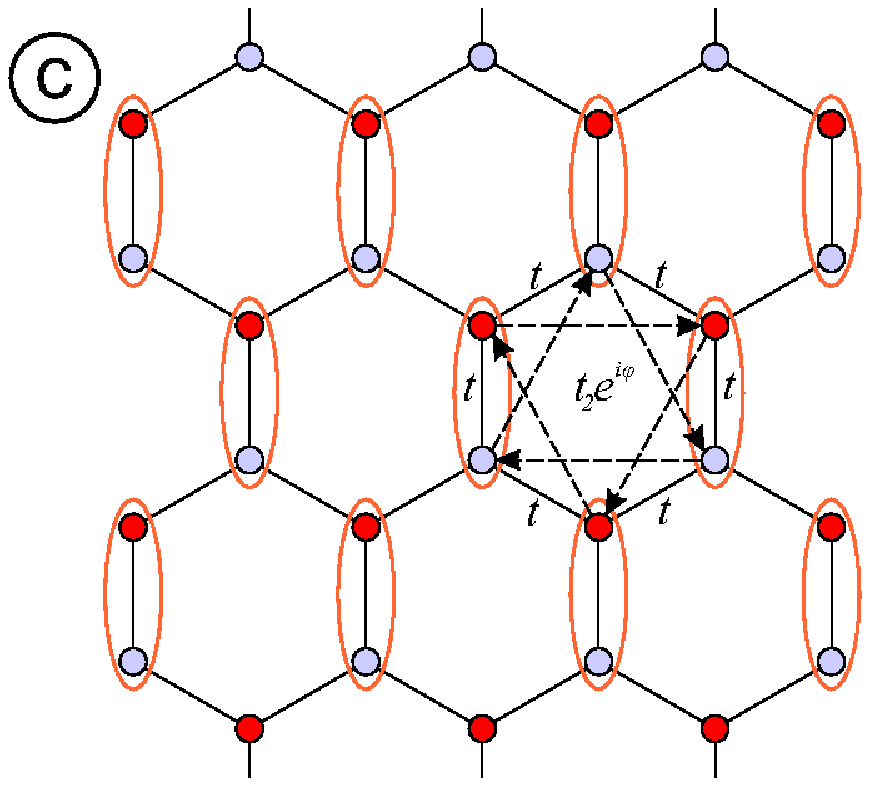}
  \caption{
      a)
      Square decorated lattice for spin subsystem:
      the four types of links (bonds) are labelled with $x$, $y$, $z$ and $xy$.
      The unit cell is a dashed square with eight sites.
      There are two types of sites: four full circles per unit cell in the center and four color ones at the border.
      Two sites with the different color placed in neighboring cells form a dimer, highlighted by a brown ellipse.
      b)
      Transformed decorated octogon-square lattice with horizontal dimers being rotated through $\frac\pi2$.
      A dashed honeycomb framework is placed behind the lattice.
      c)
      Honeycomb lattice for fermion subsystem:
      only color sites are involved, the dimer is a unit cell.
      The near-neighbor hopping $t$ between sites of different color is designated by unbroken lines,
      the next-near-neighbor hopping $t_2 e^{i\phi}$ between sites of the same color is designated by dashed lines.
      Due to the complex hopping magnitude, the direction is designated by an arrow;
      the opposite direction has the conjugated value $t_2 e^{-i\phi}$.
      A sign of the phase is represented by $\nu_{ij}$ in (\ref{eq:Hf}).
  }
  \label{Fig1}
\end{figure}

Specifically, we focus on a 2D model of a spin subsystem interacting via a Kitaev interaction with a fermion subsystem.
We look at the square-octagon variant of the Kitaev model~\cite{4} with a \spinhalf on each site (see Fig.~1a)~\cite{k2}.
The~total Hamiltonian of the present model ${\cal H}={\cal H}_{s}+{\cal H}_{f}+{\cal H}_{int}$
consists of a spin sublattice ${\cal H}_s$, a spinless fermion sublattice ${\cal H}_f$ and
an interaction ${\cal H}_{int}$ between them.
The Hamiltonian ${\cal H}_{s}$ of spin subsystem on a square decorated lattice can be written~\cite{k2} as
\begin{eqnarray}
&&{\cal H}_{s}=\Delta_x\sum_{x\mathrm{-links}}\sigma_{i}^x\sigma_{j}^x +
\Delta_y\sum_{y\mathrm{-links}} \sigma_{i}^y \sigma_{j}^y+\nonumber\\&&
+\Delta_z\sum_{z\mathrm{-links}} \sigma_{i}^z \sigma_{j}^z
+J\!\!\sum_{xy\mathrm{-links}}( \sigma_{i}^x\sigma_{j}^x+ \rho \sigma_{i}^y
\sigma_{j}^y),
\label{eq:Hs}
\end{eqnarray}
where $i$ is the nearest neighbor site of $j$ connected by
\mbox{$x$-,} $y$-, $z$- and $xy$-links as it is shown in Fig.~1a,
$\sigma_{j}^\gamma$ are the three Pauli operators at a site $j$,
$\Delta_{\gamma}$ is an exchange integral along
$\gamma=x,y,z$ direction, $J$  defines an exchange
interaction along  $xy$-links and $\rho =\pm 1$. The interaction between unit cells is defined
by the exchange interaction $J$ between spins within dimers (see Fig.~1a).

Consider a topological equivalent of the square decorated lattice, illustrated in Fig.~1b.
It is obtained by rotating the dimers that connect two neighbor cells $(u,v)$ and $(u+1,v)$ through $\frac\pi2$
so that the colored sites form a hexagonal lattice where neighbor (in terms of the hexagonal lattice) sites have different colors.
Due to topological equivalence, the first term of Hamiltonian (\ref{eq:Hs}) might be applied with the same band and mode structure.
The hexagonal lattice of colored sites provides us to introduce a second term ${\cal H}_f$ as a topological Haldane insulator state \cite{Hal} (see Fig~1c)
\begin{equation}
  {\cal H}_{f}= -t\!\sum_{<ij>}\!\! a^\dagger_{i}a_{j} - t_2 \hspace{-0.5em} \sum_{<<ij>>} \hspace{-0.5em} \exp(i \nu_{ij} \phi) a^\dagger_{i}a_{j},
  \label{eq:Hf}
\end{equation}
where $a_{j} $ and $a_{j}^\dagger$  are the spinless fermion operators with the usual anticommutation relations.
The first term represents real nearest-neighbor hopping with an amplitude $t$. The second term is the next-nearest-neighbor hopping with a hopping parameter $t_2$ and the Peierls phase $\phi$, $\nu_{ij} = \pm 1$ for the clockwise (anticlockwise) next-nearest-neighbor hopping.

The third Hamiltonian term ${\cal H}_{int}$ is governed by the interaction along $xy$-links between itinerant states of spinless fermions and
localized spins on each dimer (see Fig.~1b)
\begin{eqnarray}
&&{\cal H}_{int}=-\frac{\tau}{2}\sum_{xy\mathrm{-links}}(a^\dagger_{i}a_{j}+
a^\dagger_{j}a_{i})(\sigma^z_{i}+\rho \sigma^z_{j})+\nonumber \\
&&+ \frac{M}{2}\sum_{xy\mathrm{-links}}(a^\dagger_{i}a_{i}-
a^\dagger_{j}a_{j})(\sigma^z_{i}+\rho \sigma^z_{j}),
\label{eq:Hint}
\end{eqnarray}
where summation runs over nearest-neighbor sites $i$ and $j$ connected by a $xy$-link, $\tau$ and $M$ are the coupling parameters.

The Hamiltonian~(\ref{eq:Hs}) defines an exactly solvable \spinhalf model on the 2D square decorated lattice.
The one-particle fermion Hamiltonian~(\ref{eq:Hf}) describes free spinless fermions on the honeycomb lattice.
The Haldane model describes three phases: a metallic, topological trivial and non-trivial insulator phases with the Chern number respectively equal to 0 and $\pm 1$.
The entire model~(\ref{eq:Hs},\ref{eq:Hf},\ref{eq:Hint}) also has an exact solution in spite of the additional interaction between spin and fermion sublattices.
The $z$-components $\frac12(\sigma^z_{i}+\rho \sigma^z_{j})$ of the spin operators on dimers in~(\ref{eq:Hint}) commute with the total Hamiltonian, hence they are constants of the motion with eigenvalues $0,\pm 1$ and the interaction (\ref{eq:Hint}) can be reduced to an additional one-particle interaction between fermions along $xy$-links. The operators $\frac12(\sigma^z_{i}+\rho \sigma^z_{j})$ can be replaced with their eigenvalues: 0 realizes noninteracting spin and fermion sublattices on corresponding dimer, the $\pm 1$ case corresponds to the interacting sublattices.
First of all, topological quantum phases realized in the noninteracting spin and fermion subsystems will be considered.
We will show that the exact ground state of the model is defined by the trivial and non-trivial topological quantum phases of noninteracting spin and fermion subsystems.

\section{Spin sublattice}

The Hamiltonian of the spin sublattice can be exactly diagonalized using the representation of the Pauli operators in terms of a
related set of the Majorana fermions $b_j^\gamma$, $c_j$ ($\gamma=x,y,z$)~\cite{Kitaev}.
The site index is denoted by $j=\{s,\beta\}$, where $\beta = \overline{1,\dots,8}$ labels eight sites within a unit cell (see Fig.~1a).
We also introduce the following notation for the cells on the lattice
$s=(u,v)$, where $u$ and $v$ enumerate cells along $x$- and $y$-directions of the square lattice
($u,v=\overline{1,\ldots,n}$ where $N=n^2$ is a total number of cells).
Two cells $s$ and $s'$ are the nearest neighbors when $s=(u,v)$ implies $s'=(u\pm1,v)$ or $s'=(u,v\pm1)$, being denoted as $\langle s,s' \rangle$.
Every pair of this kind connects with a dimer --- two sites $\langle j,j'\rangle$ located in different cells near their mutual border:
$\{(u,v);7\}$ coupled with $\{(u+1,v);5\}$ and $\{(u,v);8\}$ coupled with $\{(u,v+1);6\}$.
The Hamiltonian~(\ref{eq:Hs}) can thus be rewritten
\begin{eqnarray}
 {\cal H}_{s}=-\frac{i}{2}\sum_{\gamma=x,y,z,xy}\sum_{s,s'}\sum_{\beta,\beta'}A_{s,\beta;s' \beta'}^\gamma c_{s,\beta} c_{s',\beta'},
 \label{eq:HsA}
\end{eqnarray}
with the interactions between nearest-neighbor spins inside cells
$A_{s,\beta,s,\beta'}^\gamma=\Delta_\gamma u_{s,\beta;s,\beta'}^\gamma$
and between the nearest-neighbors spins within dimers $\langle j,j' \rangle$ connecting neighbor cells
$A_{j;j'}^{xy} = J ( u_{j;j'}^x+ \rho u_{j;j'}^y)$
with $u_{i;j}^\gamma = -u_{j;i}^\gamma = i b_{i}^\gamma b_{j}^\gamma$.

There are the local plaquette $w$-operators and $q$-opera\-tors, and a static $\mathbb{Z}_2$ gauge field is associated with them.
The $w_s$-flux operator $w_s= -\sigma_{s,1}^z\sigma_{s,2}^z\sigma_{s,3}^z \sigma_{s,4}^z$ corresponds to the square in the cell $s$ (the sites marked by $1,2,3,4$, represented by full circles in Fig.~1a).

The $q$-operators are defined by two nearest-neighbor spins located at the nearest-neighbor unit cells or at the dimer $\langle j, j' \rangle$
(represented by red and blue circles in Fig.~1a)
$q_{j;j'}=  \frac{1}{2}(1-\rho\sigma_{j}^z\sigma_{j'}^z)$.

The operators $w$ and $q$ commute among themselves and with
the Hamiltonian~(\ref{eq:Hs}), and hence with $\cal H$. The plaquette operators are defined as a
product of the $ u_{s,\beta;s,\beta'}^{x}$ and
$u_{s,\beta;s,\beta'}^{y}$
operators around each plaquette, and the $q_{j;j'}$ operator is
defined by the operator $ u_{j;j'}^{x}+\rho u_{j;j'}^{y}$.
All the $u_{s,\beta;s',\beta'}^{\gamma }$ operators are
constants of motion with eigenvalues $\pm 1$, the variables $u_{s,\beta;s',\beta'}^{\gamma}$
are identified with static  $\mathbb{Z}_2$ gauge fields on corresponding bonds.
Each plaquette operator $w_s$ has two eigenvalues $\pm 1$, the eigenvalues of $q$-operator are equal to 0 and 1.
The Hamiltonian~(\ref{eq:HsA}) is reducible to a quadratic form.

\begin{figure}[tb]
\centering\normalsize
\begin{minipage}[b]{0.5\textwidth}
\includegraphics[width=0.49\textwidth]{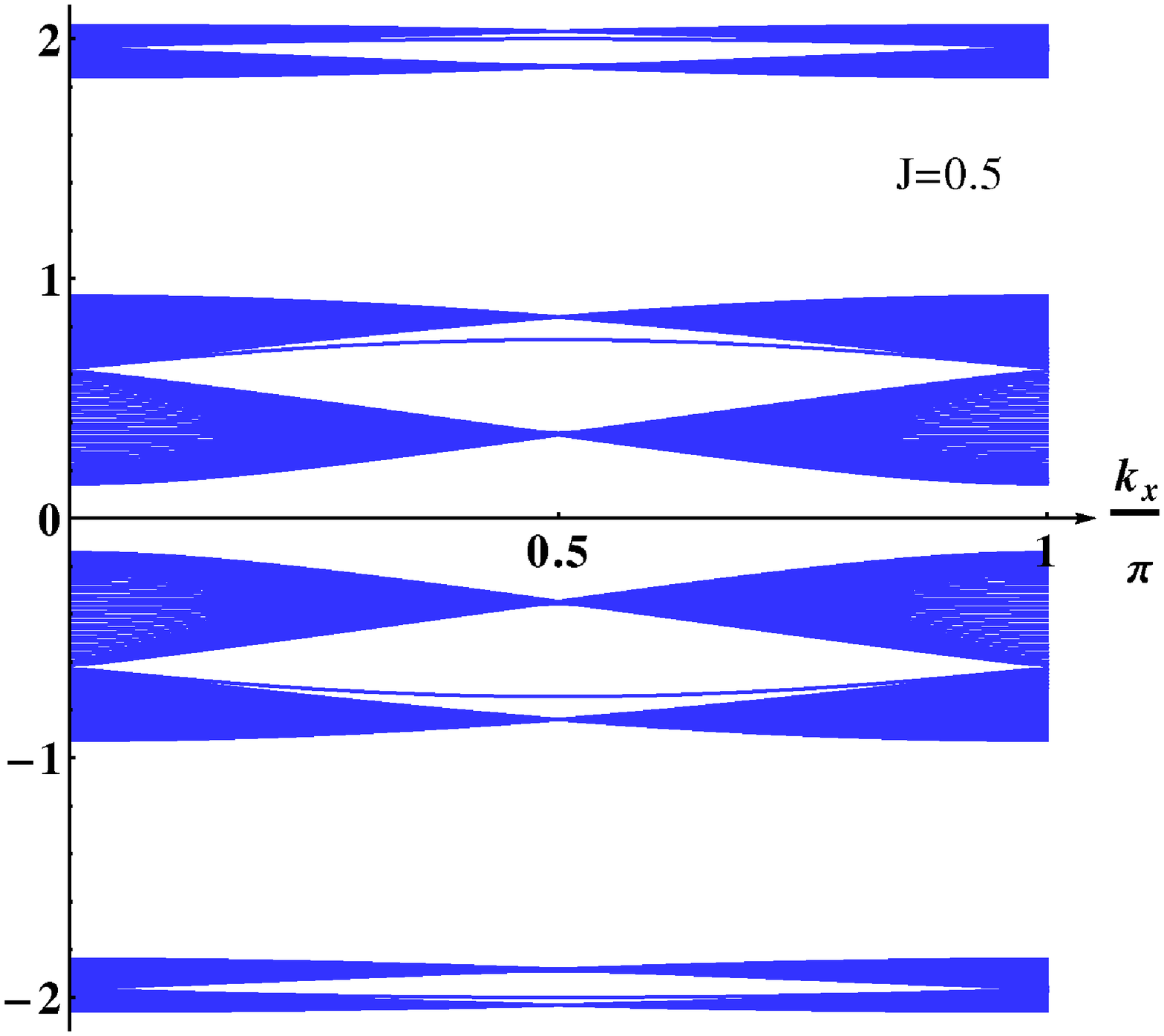}
\includegraphics[width=0.49\textwidth]{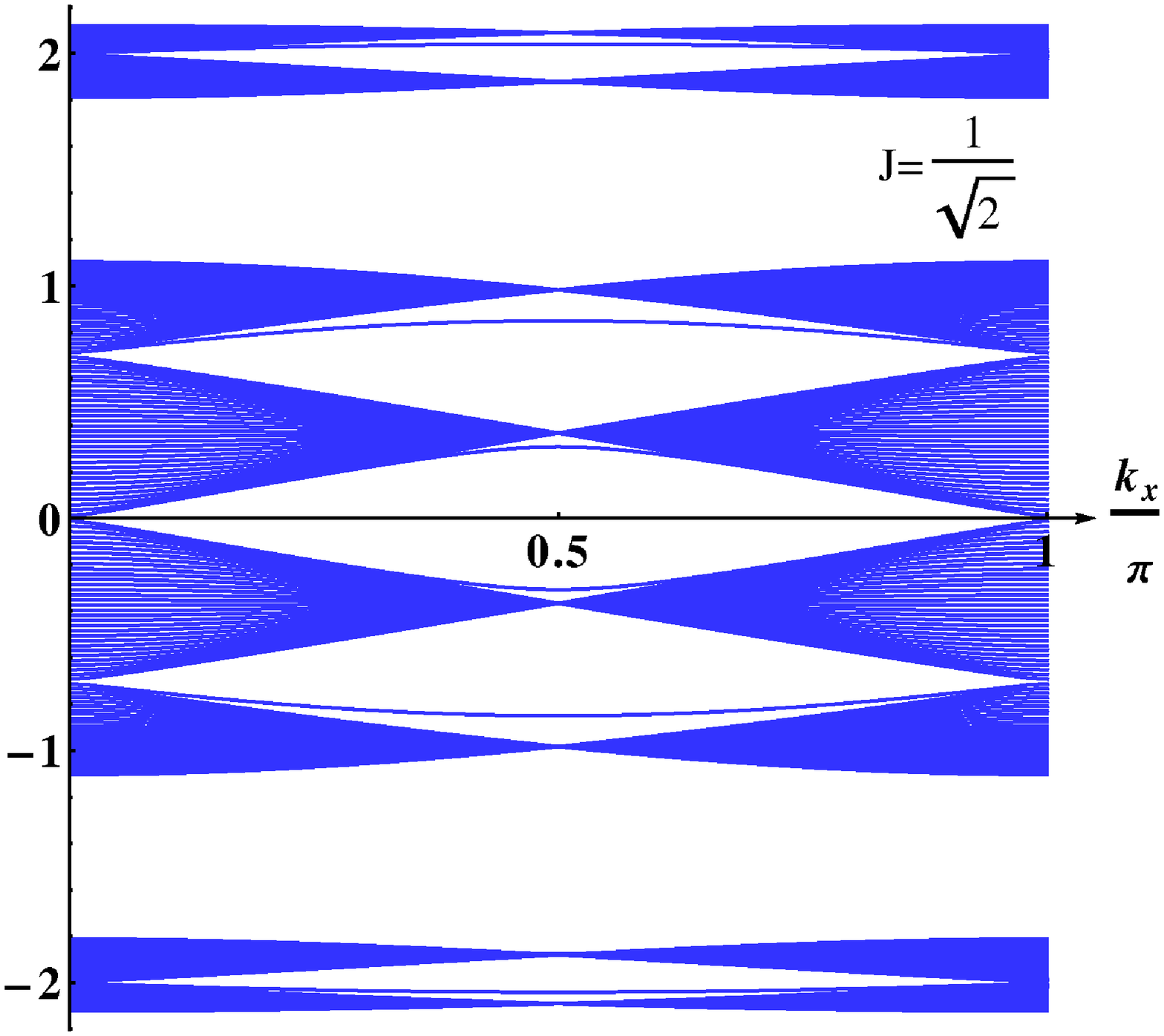}
\includegraphics[width=0.49\textwidth]{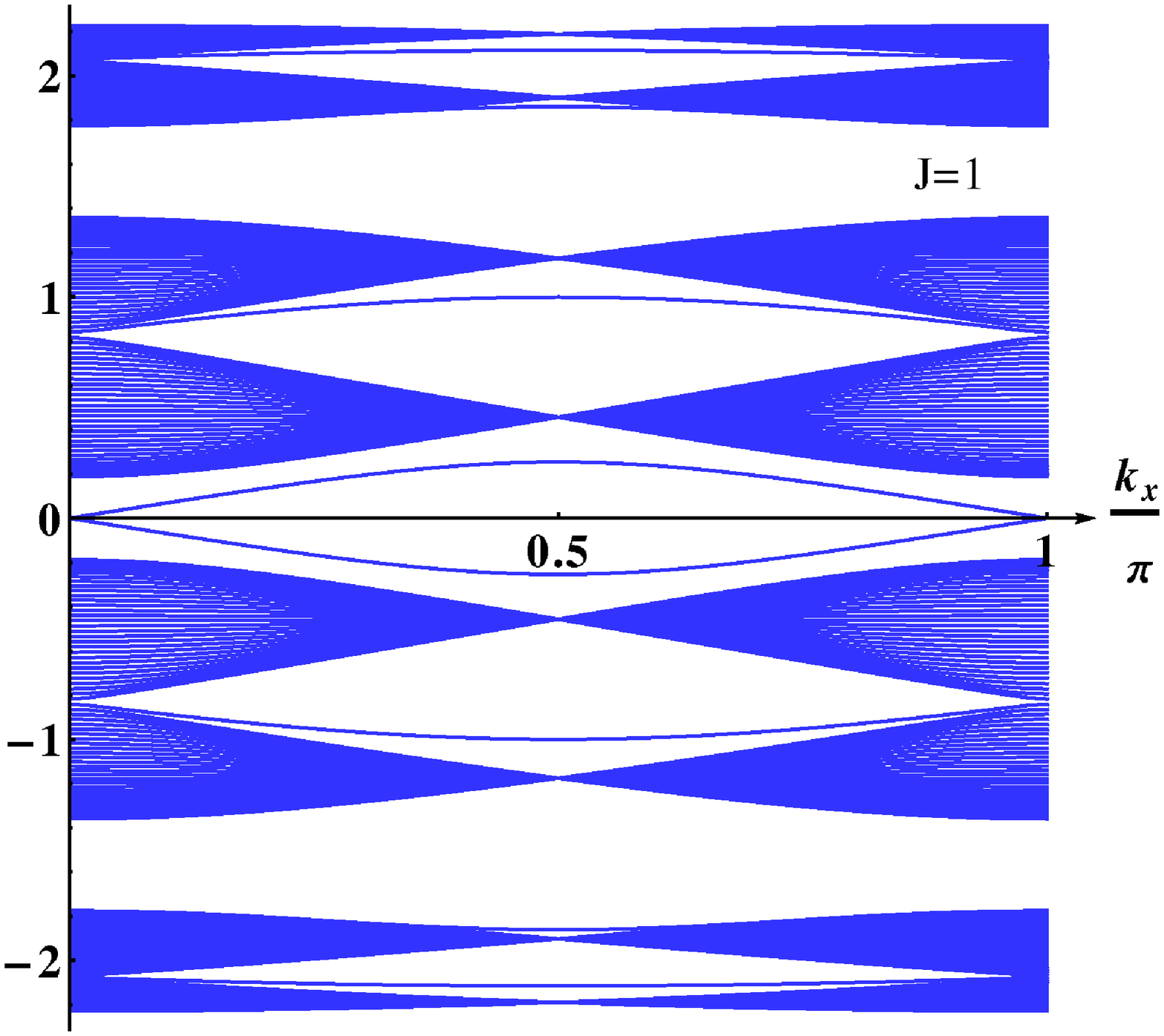}
\includegraphics[width=0.49\textwidth]{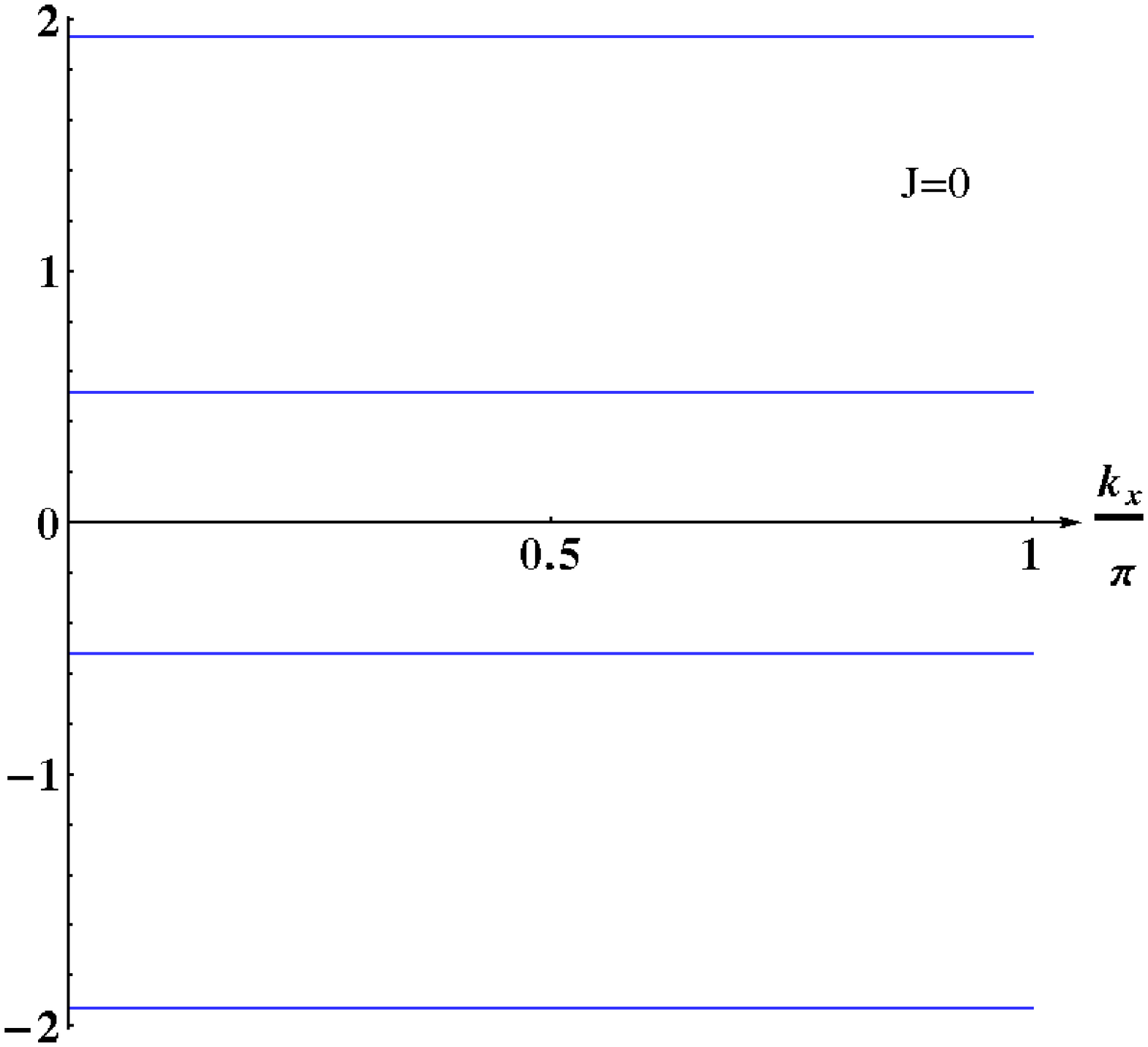}
\caption{The energy levels calculated on a cylinder with open boundary conditions in the $x$ direction as a function of
$k_x$ (in unit of $\pi$) for $\Delta_x=\Delta_y= \Delta_z=1$ and
different values of $J$: $J=1/2$ a),  $J=\frac{1}{\sqrt 2}$ b), $J=1$ c), $J=0$ or $q=0$ with twofold degeneracy d).
The Dirac point point is at $J_c=\frac{1}{\sqrt 2}$, there are gapless edge states when $J>J_c$.} \label{Fig2}
\end{minipage}
\end{figure}

To solve the model exactly, one converts each vortex sector to
free spinless fermions~\cite{Kitaev}. Numerically, we have computed the ground-state energy of
the model for a set of finite size systems and for various sets of exchange integrals.
In all cases we found out that the ground-state energy $E_{s}$ is minimized by the same uniform flux pattern.
According to Lieb's theorem~\cite{Lieb}, the energy of the free Majorana
Hamiltonian (4) is minimized when $\mathbb{Z}_2$ gauge
fields $ u_{s,i;s,j}$ are such that each $s$-plaquette has a
$\pi$-flux $\prod_{(i,j)\in s} u_{s,i;s,j}=-1$.
The vortex uniform sector with $\pi$-flux eigenvalues of the all plaquette operators $w_s$ and the same eigenvalues $q=1$ of the $q$-operators is the ground state of the spin subsystem (\ref{eq:Hs}) and (\ref{eq:HsA}).
We calculate the ground state energy of the Hamiltonian ${\cal H}_s$ for uniform $\pi$-flux sectors with $w_s=-1$ and uniform $q$-sectors with
$q_{(u,v),5;(u+1,v),7}=q_{(u,v),6;(u,v+1),8}=q$ for arbitrary values of the exchange integrals.

The model is solved analytically for the uniform configurations,
due to the translational invariance of the lattice along the unit
direction vectors $n_x=(1,0)$, $n_y=(0,1)$.  There are eight
solutions for the single particle spectrum $\epsilon_j^s(\mathbf{k})$ corresponding to eight sites per unit (see Fig.1a).
The equation written for $\varepsilon_j(\mathbf{k})=(\epsilon_j^s(\mathbf{k}))^2$ has the following form \cite{k2}
\begin{eqnarray*}
  &&\Delta_z^8 - 4\Delta_z^6\varepsilon_j (\mathbf{k}) - 2\Delta_z^4 \varepsilon_j (\mathbf{k})[\delta^2 +(Jq)^2 - 3\varepsilon_j (\mathbf{k})] -
\nonumber \\&&
    - 4\Delta_z^2 \varepsilon_j (\mathbf{k})[(Jq)^2 -\varepsilon_j (\mathbf{k})][\delta^2 - \varepsilon_j(\mathbf{k})] +
\nonumber \\&&
    + \left[ (Jq)^2 -\varepsilon_j (\mathbf{k}) \right]^2 \left[ \delta^2-\varepsilon_j (\mathbf{k}) \right]^2 +
\nonumber \\&&
     + 2\Delta_z^4 (Jq)^2 \left[ \Delta_x^2\cos (k_x+k_y) +\Delta_y ^2\cos (k_x-k_y) \right] = 0,
\end{eqnarray*}
where $\delta^2=\Delta_x^2+\Delta_y^2$, $\mathbf{k}=(k_x,k_y)$ is a wave vector.

The spectrum of Majorana fermions is always gapped, except when $(k_x,k_y)=(\pi,0),(0,\pi)$ and $(J_c \delta q)^2=\Delta_z^4$. At $J=J_c$ the spectrum is gapless with Dirac-cones. The Dirac point which occurs at $J_c$ separates two fully gapped distinct phases. The bulk gap vanishes and an edge mode
disperses through zero energy at $J>J_c$. This is illustrated in Figs~2, in which the spectrum of a relatively large cylinder is calculated for an isotropic exchange interaction between spins  at $\Delta_\gamma=1$ and different values of $J\in\{1/2,1/{\sqrt 2}, 1\}$. The spectrum
is plotted as a function of the wave vector along the boundary. At $J>J_c$ there is one edge state crossing with zero energy as it is shown in Fig~2c.
Such behavior of the spin subsystem is realized at uniform $q$-sectors with eigenvalues equal to 1.
In both uniform $q$-sectors the spin subsystem is topologically trivial with the Chern number equal to zero.

\section{Sublattice of spinless fermions}

First of all we consider topological phase transitions in the framework of the  Haldane model without a local staggered potential ($M=0$) in (\ref{eq:Hf}).
The ground-state phase diagram is defined by $t, t_2, \phi$ parameters or two parameters $t$ and $\phi$ setting $t_2=1$. The ground-state phase diagram describes phase transitions between metallic and insulator phases. The $t-\phi$ phase diagram is plotted  in Fig.~3. The lines denote the gap-closing transition between metallic and topological insulator phases, the grey regions in Fig.~3 correspond to a metallic  phase. The insulator phase is a non-trivial topological insulator phase with the Chern number $C$ equal to $1$ (at $0<\phi<\pi$) and $-1$ (at $-\pi<\phi<0$).

\begin{figure}[tbp]
\centering{\leavevmode}
\includegraphics[width=0.4\textwidth]{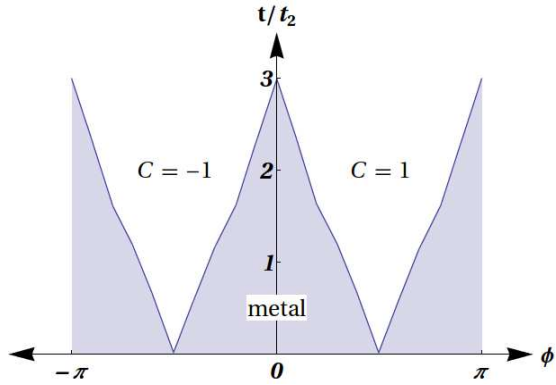}
\caption{The ground-state phase diagram of the fermion sublattice in the coordinates $\phi$, $t$ (in units of $t_2$): the solid curves correspond to quantum phase transition between metallic and topological insulator phases, the grey regions correspond to a metallic phase. } \label{Fig3}
\end{figure}

\begin{figure}[tbp]
\centering{\leavevmode}
\includegraphics[width=4.2cm]{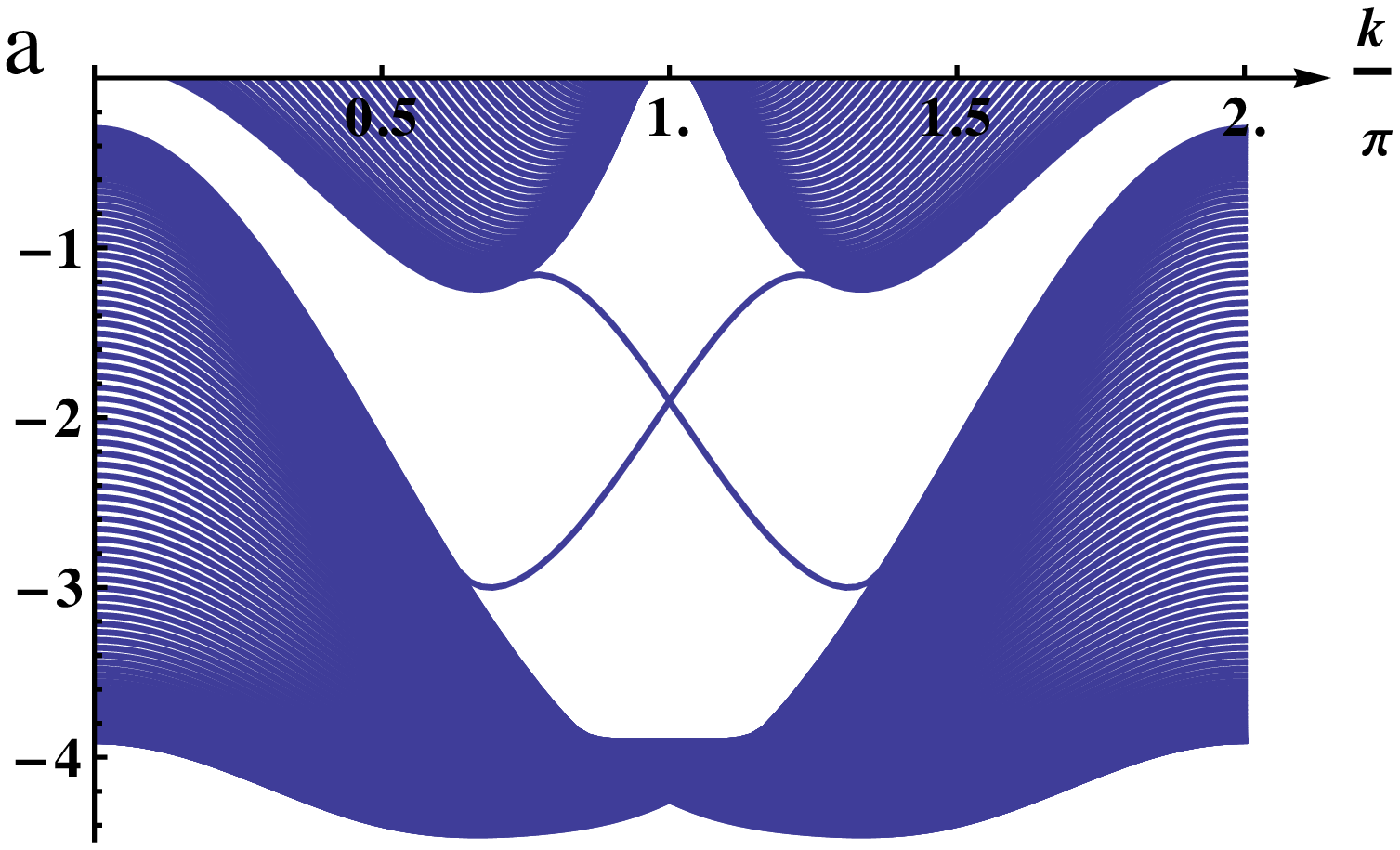}
\includegraphics[width=4.2cm]{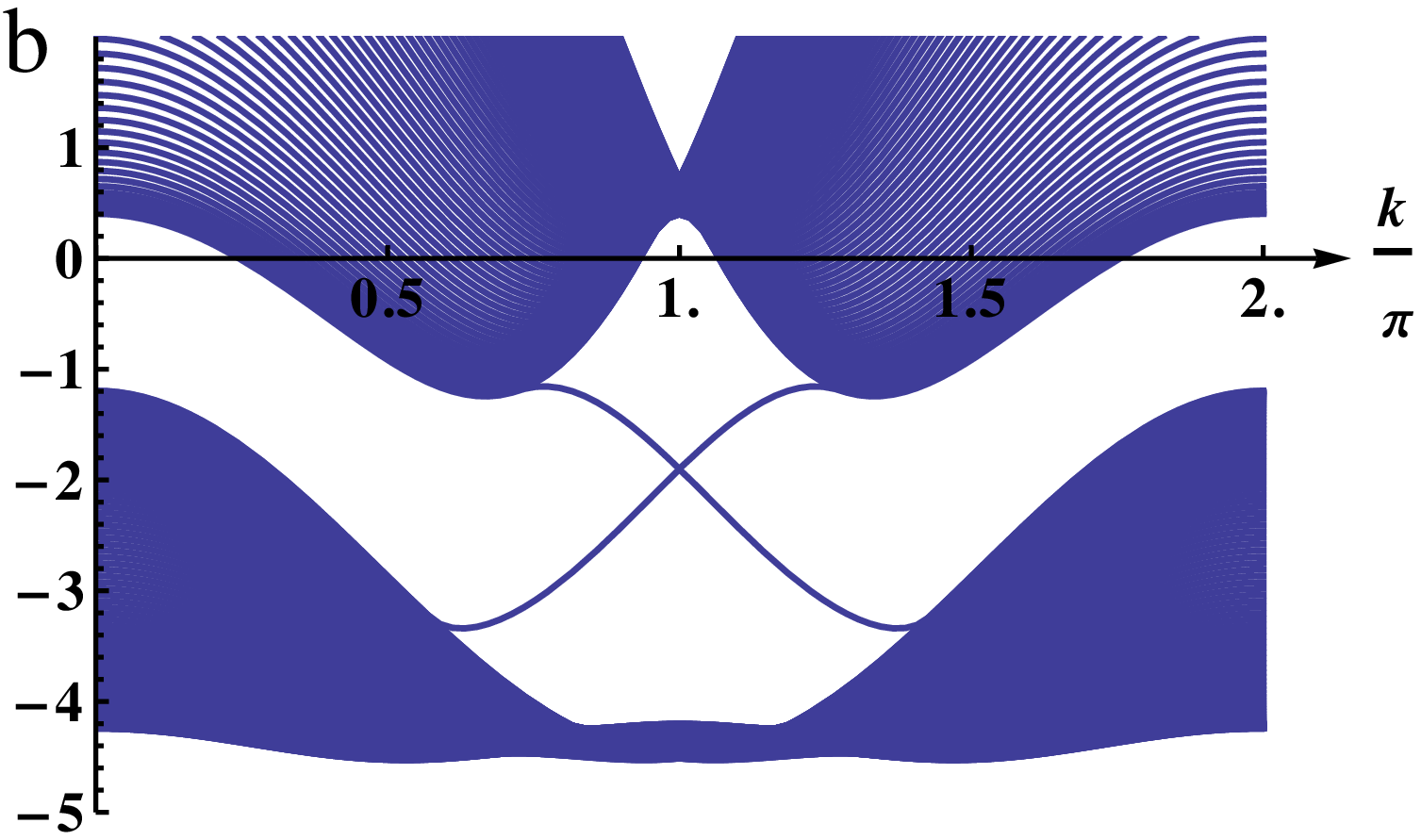}
\includegraphics[width=7cm]{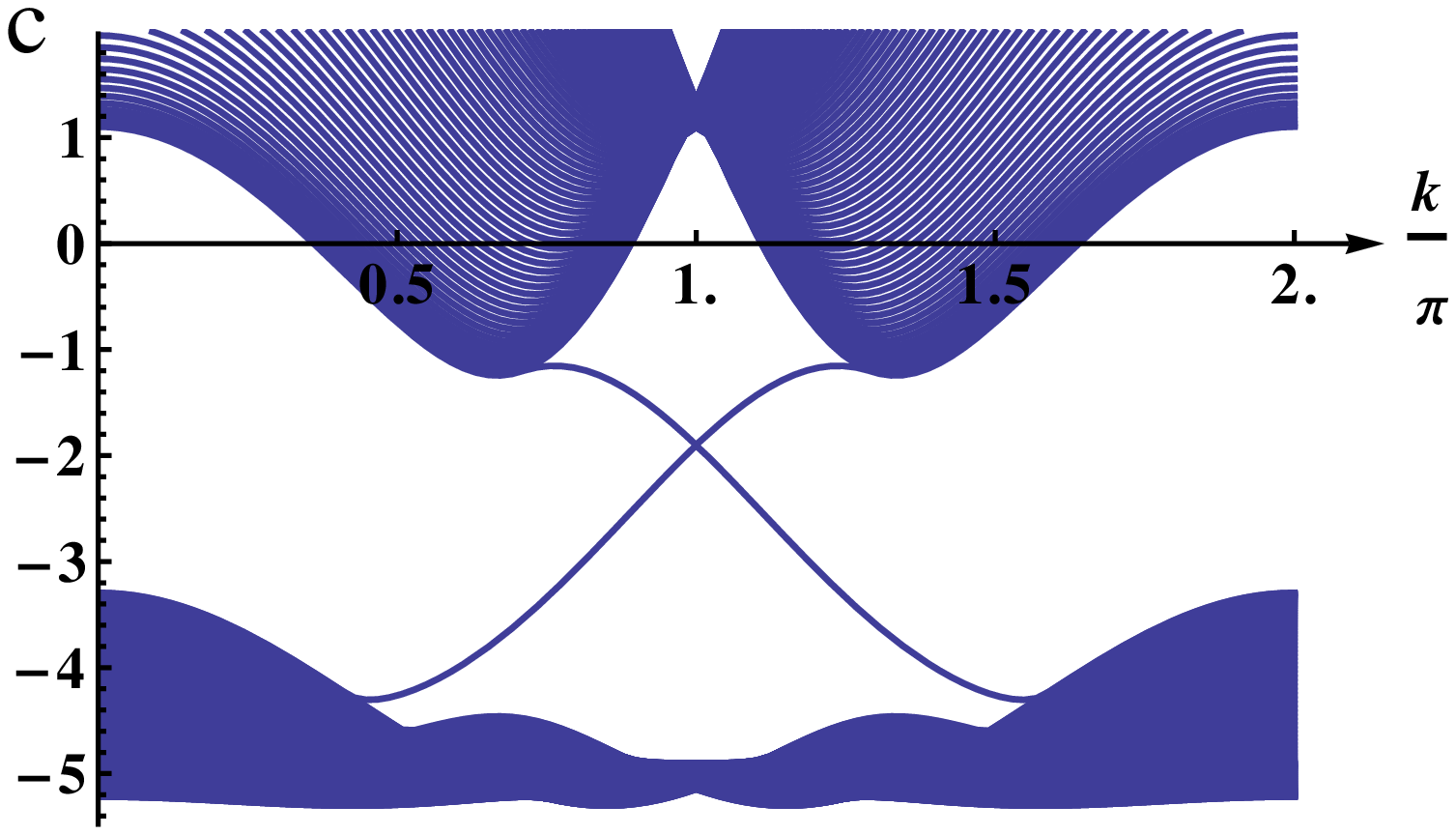}
\caption{The low energy levels in units of $t_2$ calculated for $\phi=\pi/10$ at metallic state $t=2$ a), bad metal state $t=2.3$ b) and topological insulator state $t=3$ c) on a cylinder with open boundary conditions along $(1/2,\sqrt{3}/2)$ direction as function of the wave vector $k$ in this direction. There is one edge mode connecting different subbands, it corresponds to Chern number $+1$ of insulator. Metallic state has also an edge mode, hence it might be called topological metal.} \label{Fig4}
\end{figure}

The phase of topological insulator is realized when the next-nearest-neighbor hopping $t_2$ is smaller than the nearest-neighbor hopping $t$. According to the simulations, \mbox{$t_2<3t$} is a sufficient condition for topological phase at arbitrary values of $ \phi$ (see Fig.~3). At special points $\phi=\pm\pi/2$  the topological state is realized for arbitrary values of $t$. In Fig.~4 we have plotted the spectrum  of the topological insulator with its edge states for $t=3t_2$ and $t=10t_2$.
We see that there is exactly one edge state which crosses the zero energy at the point $k=\pi$.  Thus, there exists a quantum phase transition between the metallic state and the state of topological insulator, where the bulk gap vanishes and the edge states reconnect in the insulator phase (see  in Figs~4).

When $t$ is small enough, namely the value of $t$ belongs to a grey region in Fig.~3, energy bands (dispersion surfaces projected onto energy axes) intersect due to the maximum of lower surfaces having a higher energy value than the minimum of the higher surface. In this case the Fermi level intersects with both surfaces and provokes the fermion subsystem to enter a metallic state. Nevertheless, the finite-size sample has its edge mode connecting different subbands (see Fig.~4a), this state is designated as topological metal.

\section{Phase transitions}

The complex system with the Hamiltonian ${\cal H}$ includes two phases which are defined by the uniform $q$-sector with its eigenvalues equal to 1 and 0. The first phase $q=1$ is considered above in detail. It consists of two decoupled subsystems: the spin subsystem and the topological insulator of fermion subsystem.
The Hamiltonian and the ground-state energy are respectively defined by ${\cal H}={\cal H}_s +{\cal H}_f$ and  $E_I(\Delta_x,\Delta_y,\Delta_z,J;t,t_2,\phi)=E_{s}(\Delta_x,\Delta_y,\Delta_z,J)+E_f(t,t_2,\phi)$.

Now we focus on the second phase (the uniform $q$-sector with eigenvalues $q=0$)
and calculate the ground state energy of the model 
$E_{I\!I}(\Delta_x,\Delta_y,\Delta_z;t,t_2,\phi, M, \tau) = E_{s}(\Delta_x,\Delta_y,\Delta_z,J=0) + E_f(t,t_2,\phi,M,\tau)$.
Spin cells are connected with the exchange integral $J$.
When $J=0$, the spin subsystem is a set of noninteracting cells with the total energy
$E_{s}(\Delta_x,\Delta_y,\Delta_z, J=0)=N\varepsilon(\Delta_x,\Delta_y,\Delta_z)$,
where $\varepsilon(\Delta_x,\Delta_y,\Delta_z)$ is an energy, calculated for each cell separately,
consisting of two twofold degenerated flat levels (see Fig~2d).
Comparing $E_{I}$ and $E_{I\!I}$ energies, we calculate the ground-state energy and the ground-state phase diagram of the proposed model.

In order to calculate the term $E_{f}$, we consider the fermion subsystem with the Hamiltonian ${\cal H}_f+{\cal H}_{int}$,
where all the operators $\frac{1}{2}(\sigma_i^z+\rho\sigma_j^z)$ are replaced with their eigenvalues $\pm 1$ uniformly in the term~(\ref{eq:Hint}).
At $\tau=0$ the fermion subsystem is reduced to the original Haldane model,
which is similar to the model considered in the previous section, but with additional staggered potential with the amplitude $M$ (the last term in~(\ref{eq:Hint})).
In the $M-\phi$ phase diagram of the Haldane model, the line $M/t_2=\pm 3\sqrt{3}\sin \phi$ of the phase transition between topological trivial (with the Chern number zero) and non-trivial (with the Chern number $\pm 1$) phases separates different insulating states.
This phase diagram is plotted in Fig.~5 in the coordinates $(\phi,M,t)$ for $\tau=0$.

\begin{figure}[tbhp]
\centering{\leavevmode}
\includegraphics[width=2.in]{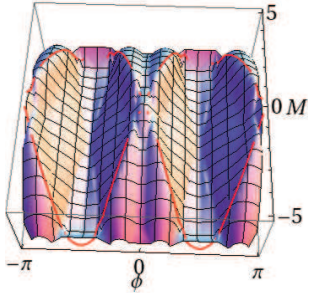}
\caption{The phase diagram of the Haldane model in the coordinates $(\phi, M,t)$ where $t$ is perpendicular to plane $(\phi, M)$ and $t_2=1$.
The line  $M=\pm 3\sqrt{3}\sin \varphi$ arranges on the `ridge', the `mountains' is the metal phase. } \label{Fig5}
\end{figure}

\begin{figure}[tbhp]
\begin{minipage}[b]{0.5\textwidth}
\centering{\leavevmode}
\includegraphics[width=\textwidth]{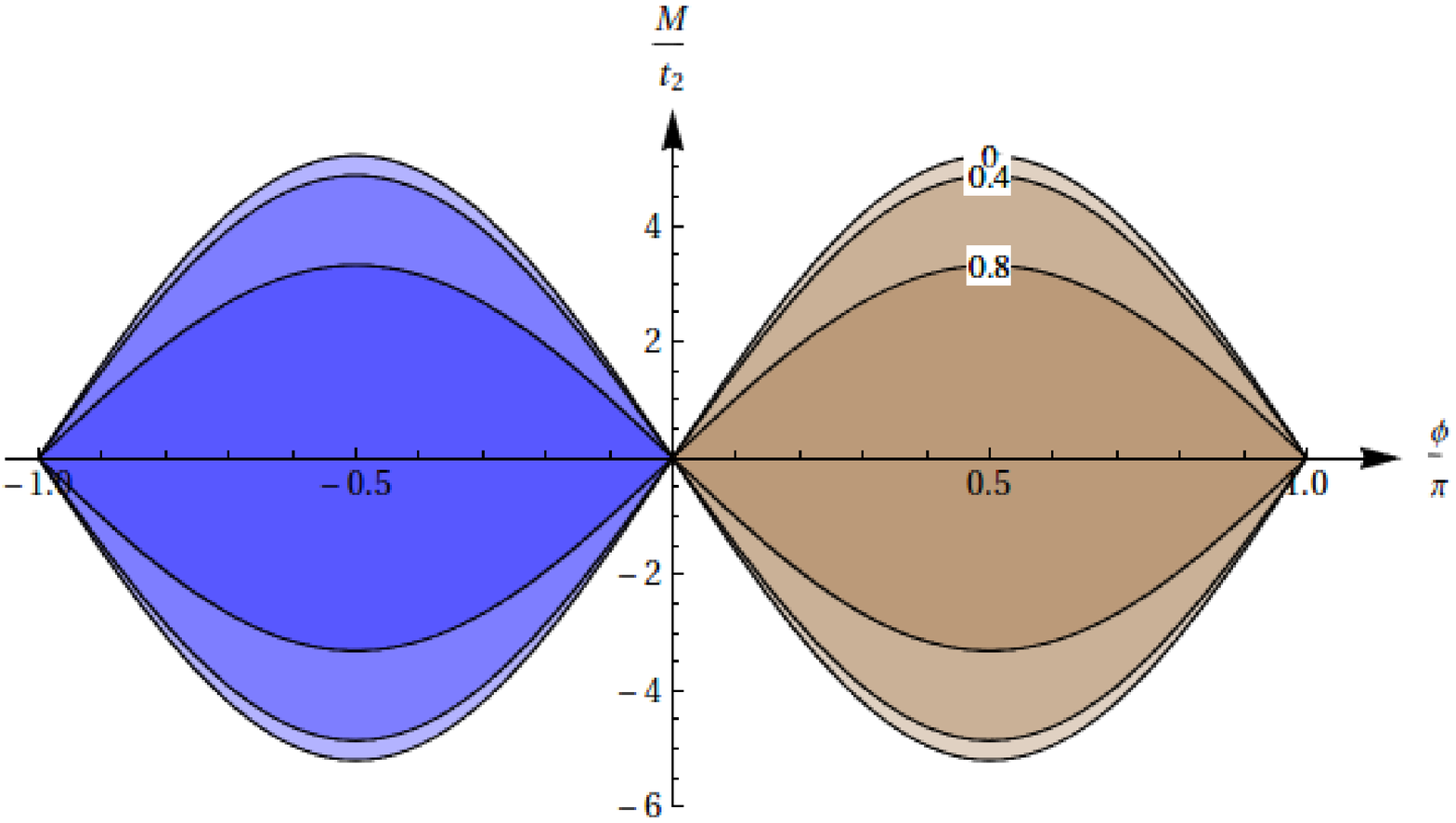}
\includegraphics[width=0.7\textwidth]{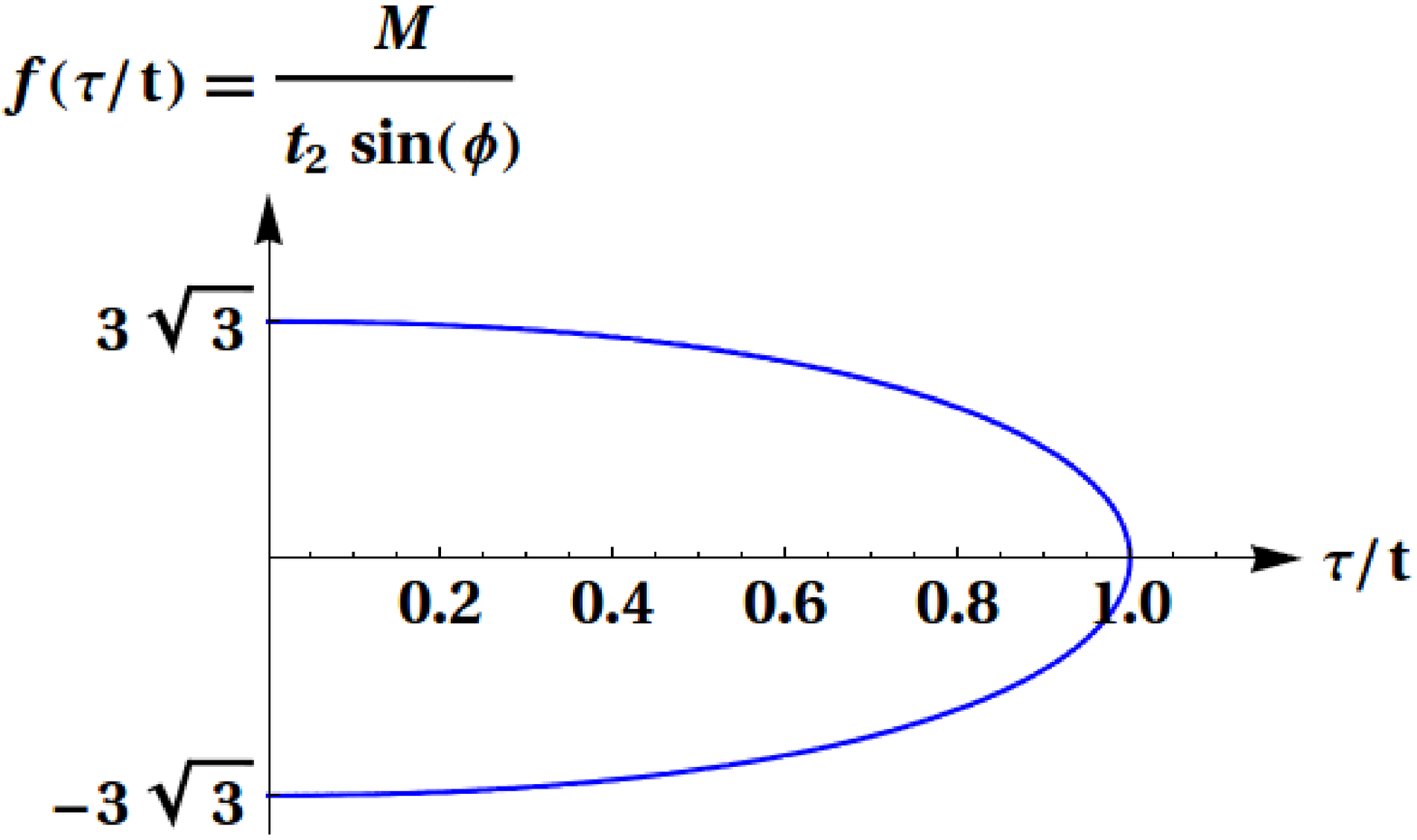}
\caption{top: Topological phase diagram of ${\cal H}_f + {\cal H}_{int}$ fermion model in II phase ($q=0$)
  in coordinates $(M/t_2, \phi)$ for different values of $\tau/t$ according to (\ref{eq-haldane}).
  The widest topological region is realized in pure Haldane model ($\tau=0$), it shrinks with an increase of $\tau$ and vanished completely when $\tau=t$.
  bottom: the dependency between the topological region's size and $\tau/t$.
} \label{Fig6}
\end{minipage}
\end{figure}

The ground-state phase diagram of the Haldane model in the coordinates $\varphi, M,t$ (at $t_2=1$ and $\tau=0$) is shown in Fig.~5, the 2D surface separates  metal and insulator phases. In this figure, the system of itinerant fermions is metallic below the surface ($t < t_m(M,\phi)$ in units of $t_2$).
Above the surface the system is an insulator, topologically trivial or non-trivial. The sufficient condition of the insulating state is $t>3t_2$, as it was stated before.
It was proved mathematically that the shape of the metal-insulator transition surface does not depend on the value of $\tau$, or it might at most be weakly dependent.

When $\tau\neq 0$, the additional hopping within each dimer is present and the total hopping is $t\pm\tau$
while the hopping between nearest-neighbor sites of different dimers remains $t$.
Up to energy minimization, the sign is chosen for maximal value of the effective hopping of fermions into dimers,
thus hopping within dimer is $|t|+|\tau|$ and an assumption $t,\tau\geq0$ is sufficient for further consideration.
The topological phase transition occurs with an arbitrary value of~$t$, when the system traverses the Haldane curve in the $M-\phi$ diagram.
In the case $\tau\neq 0$, the line of the quantum phase transition is deformed according to the following formula
\begin{equation}
    \frac{M}{t_2 \sin \phi} = \pm(3 +\tau/t)\sqrt{3-2\tau/t-(\tau/t)^2},
    \label{eq-haldane}
\end{equation}
its form, shown in Fig.~6a, is similar to the case $\tau=0$.
If $\tau\geq t$, the fermion subsystem always stays in topologically trivial state.

\begin{figure}[tp]
\centering{\leavevmode}
\begin{minipage}[b]{0.5\textwidth}
\includegraphics[width=0.48\textwidth]{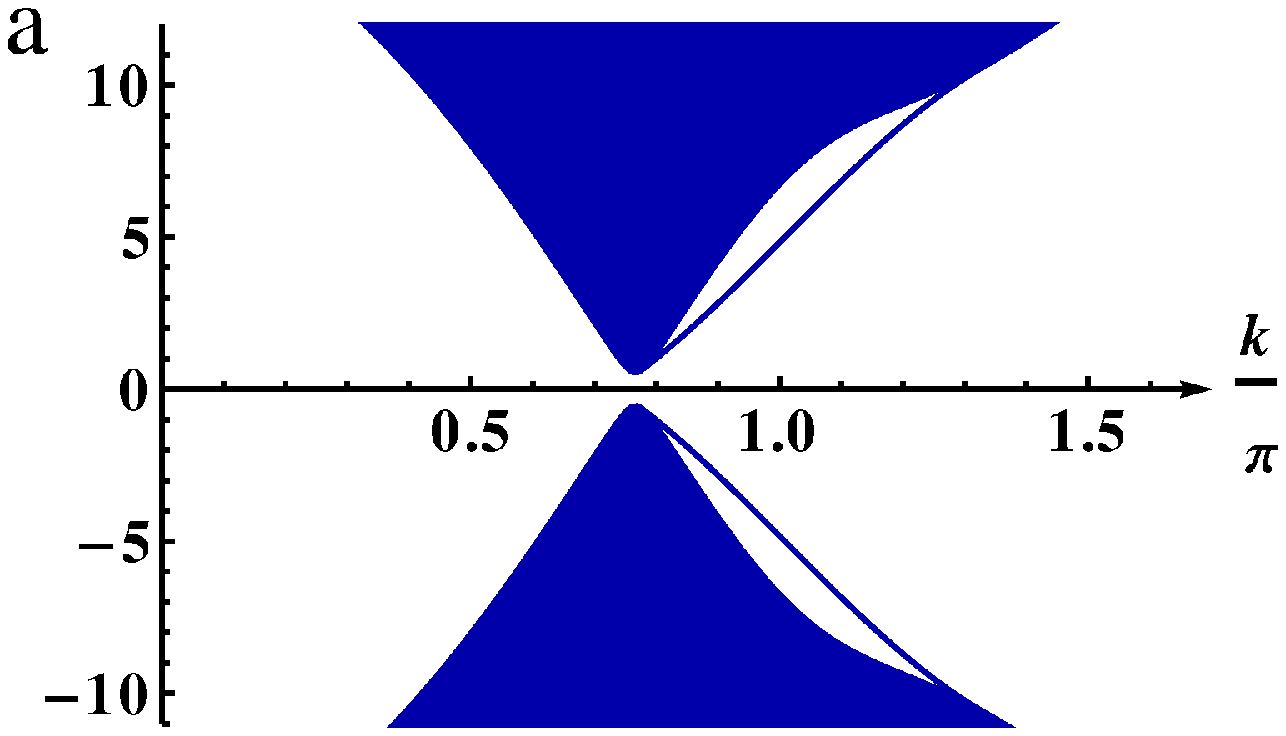}
\includegraphics[width=0.48\textwidth]{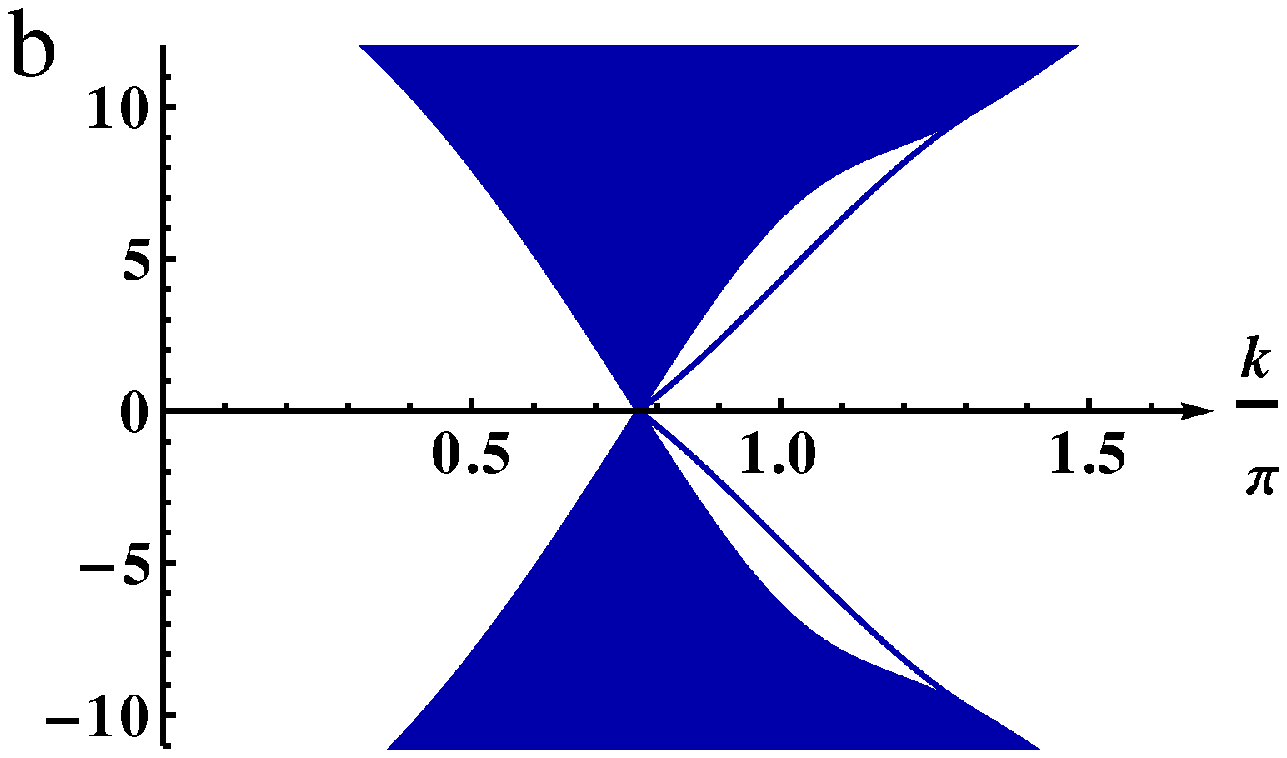}
\includegraphics[width=0.48\textwidth]{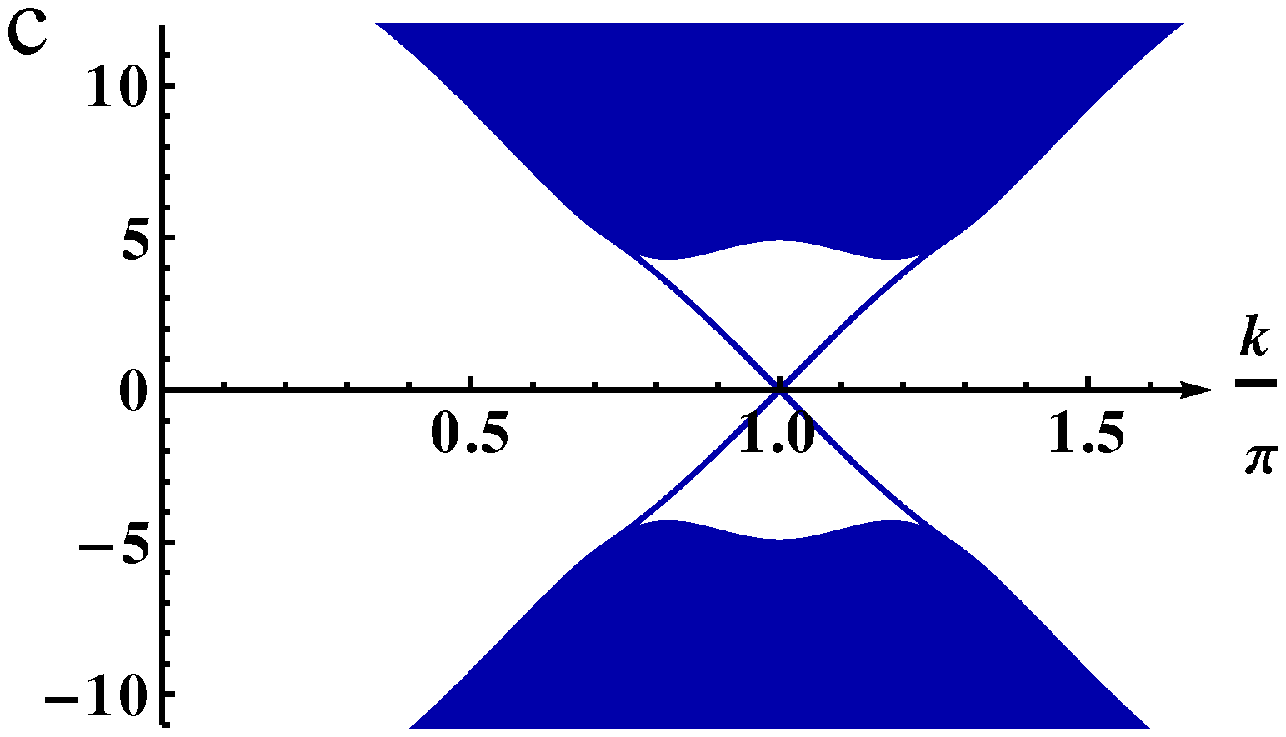}
\caption{The energy levels of spinless fermion subsystem calculated on a cylinder with open boundary conditions in the $(1/2,\sqrt{3}/2)$ direction as a function of the wave vector along this direction ($k$ in unit of~$ \pi$) for $t_2=1$, $t=10$ , $\phi=\pi/2$, $\tau=5$, and $M=3\sqrt3$ a), $M=\frac{7\sqrt 7}4$ b) and $M=0$ c).} \label{Fig7}
\end{minipage}
\end{figure}

In Figs~7 we show the low energy band structure of fermions with edge modes calculated in the framework of the Haldane model at the typical point $\phi=\pi/2$ for $t_2=1$, $t=10\gg t_2$ and $\tau=5$ and different values of $M$: $M=3\sqrt3$ (in the trivial topological phase with $C=0$), $M=\frac{7\sqrt 7}4$ (at the point of the phase transition) and $M=0$ (in non-trivial topological phase with $C=1$).

The ground state energy per cell $\varepsilon = \varepsilon_s +\varepsilon_f$
is defined by both spin and fermion sublattice energy expressed by the spectra of spin and fermion subsystems:
$$\varepsilon_s=\sum_{j}\sum_{\mathbf{k},\epsilon_j^s(\mathbf{k})<0} \epsilon_j^s(\mathbf{k}), \quad
  \varepsilon_f=\sum_{j}\sum_{\mathbf{k},\epsilon_j^f(\mathbf{k})<0} \epsilon_j^f(\mathbf{k}).$$
Let us compare the energies $\varepsilon$ of states with different eigenvalues of the $q$-operators.
 It is convenient to consider the ground-state phase diagram of the system in the coordinates $M,\tau,J$ for given values of the parameters that characterize the spin ($\Delta_\gamma$) and fermion ($t,t_2,\varphi$) subsystems.
The 2D surface separates the topological phases with different or same Chern numbers.
In the insulator phase~I the fermion subsystem is defined by the Chern numbers $\pm 1$, whereas the Chern numbers of the insulator in the phase~II are  both $C=0$ and $C=\pm 1$ depending on the parameters of the Hamiltonian.
The phase transition between I and II phases is accompanied by a fundamental rearrangement of the spectrum of the spin subsystem from band states in~I to flat-band states in~II.
Let us note that on the 2D critical surface in Fig~8 which separates phases with different topologies of fermion subsystem there are quantum phase transitions (between phases with both the same and different Chern numbers) without the gap closing.

\begin{figure}[tp]
\centering{\leavevmode}
\includegraphics[width=7cm]{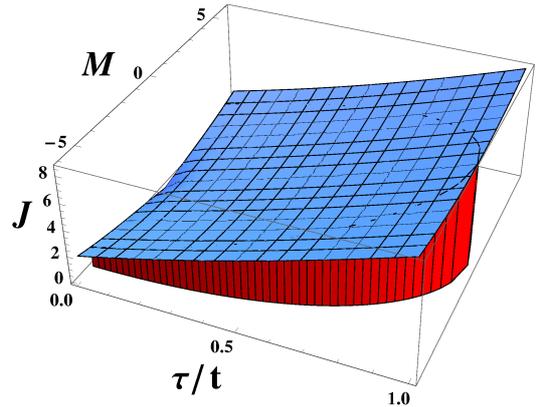}
\caption{The ground-state phase diagram in the coordinates $\tau, M,J$. The phase~I is realized above blue surface, whereas the phase~II is realized below it, in topologically trivial (outside red surface) or non-trivial (inside the region limited by surfaces) state.} \label{Fig8}
\end{figure}

\begin{figure}[tbhp]
\begin{minipage}[b]{0.5\textwidth}
\centering
\includegraphics[width=0.48\textwidth]{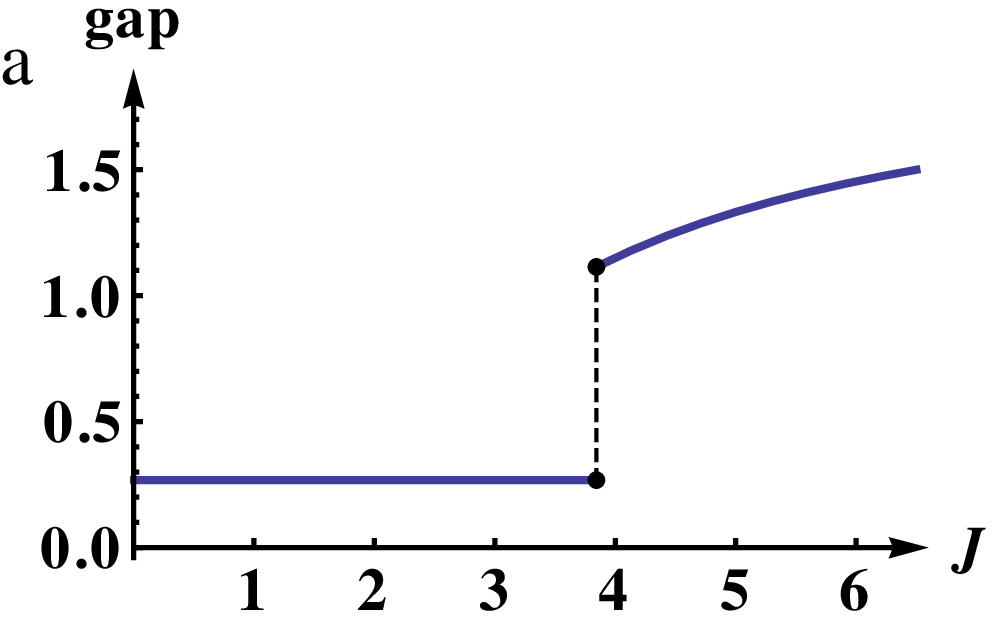}
\includegraphics[width=0.48\textwidth]{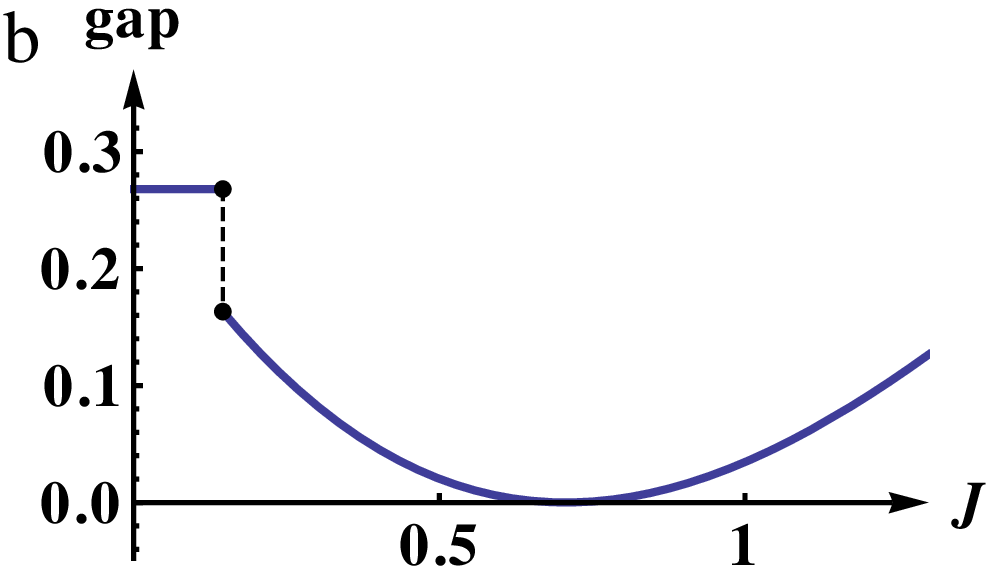}
\includegraphics[width=0.6\textwidth]{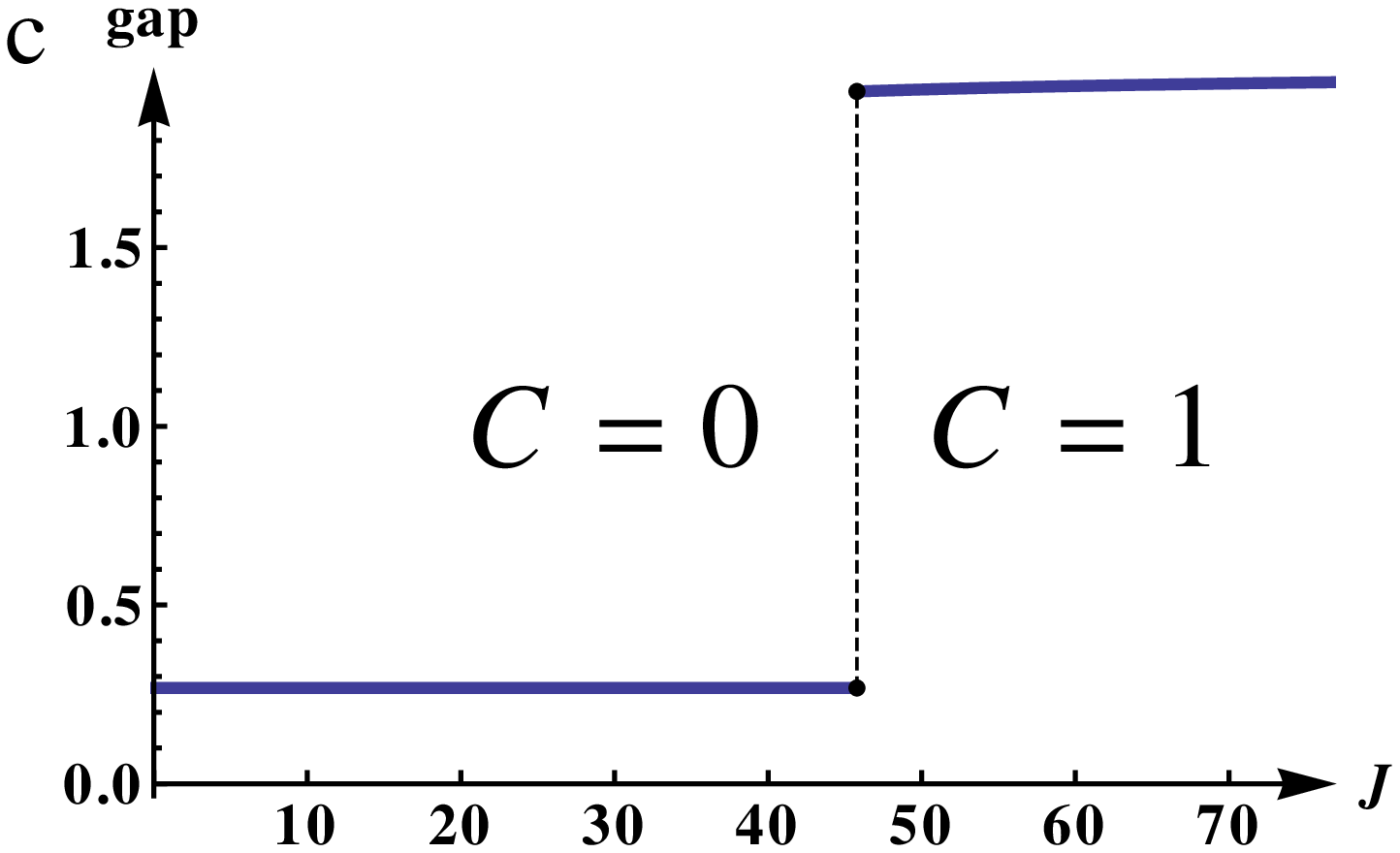}
\caption{Schematic illustration to Fig.~8: the gap depending on $J$ for different values of the fermion subsystem's parameters $M=0, \tau=0.5$ a), $M=0.5,\tau=0$ b), $M=5, \tau=0.5$ c) for $t=10$, $t_2=1$, $\phi=\frac\pi2$. The dashed line designates a gap discontinuity near the phase transition point. Both a) and b) demonstrate phase transition while preserving the Chern number $C=1$ between two topologically nontrivial phases of fermion subsystem. In the last case c) two phases have different Chern numbers.}
\end{minipage}
\end{figure}

Fig.~8 displays numerical results for definition of the regions of the stable phases I and II as function of $J$ and the parameters of
the interaction $\tau$ and $M$,  where phases I and II are stable above and below the surface, respectively.

We illustrate the transition progress in Figs~9 for the different cases.
The low energy gap is defined by the gap of the spin subsystem for chosen parameters,
because the gap of the fermion subsystem is usually wider, except the case of $M$ and $\tau$ being near the curve shown in Fig.~6 of topological phase transition of fermion subsystem, and it is still a positive constant for $t\neq 0$.
The gap is a constant in phase~II, because it does not depend on $J$.
The spectrum of Majorana fermions is gapped along a line of the phase transition.
A gap closing in Fig.~9b corresponds to Fig.~2b and does not entail a change of any symmetry.
A low energy gap has a jump from a non-dispersive flat-band state  (in the phase~II) to a dispersive band state (in the phase~I) at the point of the topological phase transition.
In Fig.~9a and Fig.~9b the system is in topological phase, as it is evidenced by the presence of gapless edge states, this phase transition occurs with different $\mathbb{Z}_2$ gauge fields without gap closing at the point of the topological phase transition.
In that case, both phases belong to the same topological class (with the same Chern number $\pm 1$).
Another type of phase transition is shown in Fig.~9c with parameters $M$ and $\tau$ chosen from the region outside of the red surface in Fig.~8 or outside colored region in Fig.~6, since it not only breaks $\mathbb{Z}_2$ symmetry, but it also changes the Chern numbers at the point of the topological phase transition. The subsystem of spinless fermions is in a topological insulator state in a region of large $J$ (phase~I) and in a trivial state when $J$ is low (phase~II).

The quantum phase transition corresponds to the change of the dimers' state,
in other words the transition from singlet (with $\frac12(\sigma^z_i+\sigma^z_j)=0$) to triplet
states (with $\frac12(\sigma^z_i+\sigma^z_j)=\pm 1$) of dimers. The $xy$-links  couple the
nearest-neighbor cells, therefore the  states of dimers
define the type of band spin excitations in the system. In the triplet
state of dimers at $\rho =-1$ the band excitations are two-particle excitations
paired on the dimers because one particle excitations are
forbidden. Another situation takes place in the singlet state of
dimers at $\rho =1$, where only one-particle excitations are realized (two-particle
excitations are forbidden). The spectrum of Majorana fermions is gapped along the surface of the phase transition.
Note that the phase transitions between topological states of the system are realized without gap closing at the point of the phase transition.
Such an unusual phenomenon can occur when the first-order of phase transition between topological states with different static $\mathbb{Z}_2$ gauge field configurations is realized.
The ground states of the system are different for the uniform flux
sector with triplet and singlet states of dimer. In the uniform sector with triplet states of dimers
at $\rho =-1$ the ground states and the full set of excited states are BCS
ground state wave functions and BCS excited wave functions over
the vacuum. The BCS wave functions can be mapped to a system of
spinless fermions with pairing of the fermions within the dimers.  The total energy of the system decreases with the increasing value of the hopping integral of fermions, the minimum of the energy is reached at the same local moments of the dimers when the amplitude of the hopping integral is equal to $\vert t \vert+\vert \tau\vert$. At $\rho=1$ the phase II is the Ising magnetic with magnetization equal to $2N$ (2 is a local spin of dimer).

\section{Concluding remarks}

In the framework of the proposed model
we have calculated the ground
state phase diagram of the topological insulator interacting with magnetic subsystem as a function of the exchange integrals,
the hopping amplitudes and the local staggered potential.
We have studied an exactly solvable model defined on the 2D square
decorated lattice that realizes quantum phase transitions  between phase states with singlet and triplet states
of dimers. This phase transition is realized
without the low energy gap closing.
As a rule, the low energy
spectrum is gapless at the point of a topological phase transition. The model exhibits a complex ground-state phase diagram. Quantum phase
transitions of both spin and fermion subsystems separate topological phase states.
The Majorana fermions localized on the links  are defined by
a static $\mathbb{Z}_2$ gauge field. A configuration
of the gauge field is changed at the point of the phase transition.
The first-order phase transition is characterized by a jump of local moment of the dimer.
The wave function of itinerant Majorana fermions defines one-particle (at $\rho=1$) and two-particle (at $\rho=-1$)
spin excitations inside the dimers.
We have shown that this phase transition is accompanied by a fundamental rearrangement of the spectrum of the spin subsystem
from band-states of the Majorana fermions to flat-band states.
We have established the quantum phase transitions between topological phases with both different values of the Chern numbers and the same Chern numbers that define the topological state of the system.
We do not discuss the question of the realization of the model proposed. Nevertheless, the family of complex topological lattice models opens new opportunities to study the interplay between topological magnetic and fermion systems in the search for new quantum phases and new quantum phase transitions.


\end{document}